\documentclass[twocolumn,aps,secnumarabic%
,showkeys,amsmath,amssymb,prb]{revtex4}

\usepackage{graphicx}
\usepackage{dcolumn}
\usepackage{bm}

\begin{document}


\title{Thermal emission by photonic micro-textured surfaces}

\author{Jones T. K. Wan}
\affiliation{Department of Physics,
Hong Kong University of Science and
Technology, Clear Water Bay, Kowloon, Hong Kong.}

\author{C. T. Chan}
\affiliation{Department of Physics,
Hong Kong University of Science and
Technology, Clear Water Bay, Kowloon, Hong Kong.}


\date{\today}

\begin{abstract}
Ordinary metallic photonic crystals (PCs) have photonic band gaps in which the
density of states (DOS) is strongly modified. Thermal emission of photons can
be suppressed and enhanced accordingly. We consider the thermal emission
characteristics of a metallic photonic crystal slab with a tunable thickness
which in the thick limit approaches that of a photonic crystal and in the thin
limit approaches that of a textured surface. We find that a thick photonic
crystal suppresses emission in a specific range while a thin slab suppresses
low frequency emission.
\end{abstract}

\keywords{photonics, thermal emission, nanostructures}
\maketitle

The study of thermal emission has be an active subject of research area
for many years.\cite{planck_1959,howell_2002}
At thermal equilibrium, the emissivity of a material at each frequency
is equal to its absorptivity, if the transmittance is zero.
This makes metals good thermal
radiators at infrared and optical frequencies as common metals are usually
good absorbers in these frequency ranges.
Recent developments of photonic crystals has opened a new arena in controlling
the absorption spectrum.\cite{hesketh_nature_1986,joannopoulos_1995,dowling_pra_1999,pendry_jpcm_1999,chan_prl_2000,chan_opt_express_2001}
If we want to suppress radiation at infrared frequencies, then
by manipulating the geometry and materials parameters, one can design
a metallic photonic crystal with large band gaps over the infrared
frequencies.
This makes the study of thermal emission by photonic crystals a
fruitful research area, which has witnessed a rapid growth in recently years.
However in reality, difficulties arise in fabricating
three-dimensional (3D) photonic crystals, which prohibits its potential
applications. 
Recently, Fleming et al.\cite{fleming_nature_2002} reported that only one
layer of photonic crystal is sufficient to achieve strong attenuation
of electromagnetic radiation. 
Regarding to their findings,
here we propose a model system
which only consists of a tungsten micro-particle layer, sitting on
a thick tungsten slab.
The present geometry exhibits excellent enhancement of absorption in
the range of optical frequencies, while suppressing the absorption
in other frequency ranges.

We begin by considering a tungsten face-centered cubic (FCC) photonic crystal
with a lattice constant $a$= 0.5 $\mu$m, for simplicity, the spherical
tungsten micro-particles are embedded in a background medium with a dielectric
$\epsilon_0$=1.  The interparticle distance $\alpha$ is given by
$\alpha=a/\sqrt{2}=0.354$ $\mu$m and the filling ratio $f$ is given by
$f=0.65$ ($r=0.169$ $\mu$m).  The absorbance ($A$), reflectance ($R$) and
transmittance ($T$) as well as the photonic band structure are calculated by
using multiple-scattering formalism.\cite{modinos_cpc_1998} It should be noted
that, the present geometry is not the only method of choice, but it allows us
to use the multiple-scattering formalism, which has be proven to have a high
accuracy and is less computational costly than the finite-different time
domain (FDTD) method.\cite{joannopoulos_prl_2004}

%
\begin{figure}[htb]
\includegraphics[height=0.9\columnwidth,angle=270]{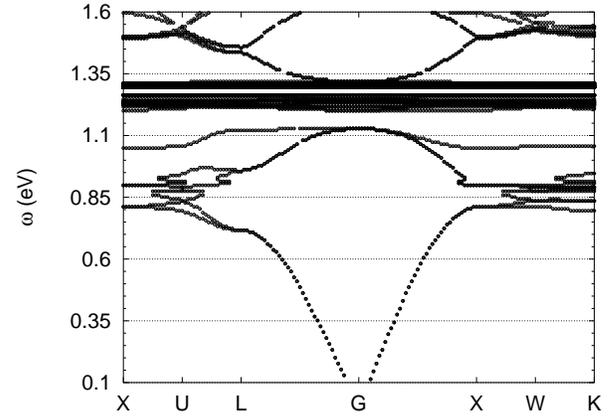}
\caption{The photonic band structure of a FCC tungsten photonic crystal,
$a$= 0.5 $\mu$m and $f=0.65$. A large band gap at infrared frequencies is
found.}
\label{fcc_band}
\end{figure}
%
In Fig.~\ref{fcc_band}, we show the photonic band structure of the FCC
tungsten photonic crystal. As seen, its band structure exhibits a large
directional band gap at infrared frequencies, which suggests that the tungsten
photonic crystal can effectively suppress the infrared radiation.  However, it
should be noted that the band structure calculation was based on an infinite
lattice structure, while a realistic PC is constructed by a finite number of
particle layers.  Here one may ask {\it how many} layers are sufficient for
the purpose of thermal emission?  And more importantly, is it possible to use
just {\it one} layer of (metallic) particles?  In order to explore the answer
to these questions, we consider a model system which consists of a single
(111) layer of tungsten particles and a tungsten slab only. The tungsten slab
is acting as a substrate and it also ensures the overall transmittance is zero
at all frequencies.  The tungsten slab is put at a distance $l$ behind the
center of the tungsten particles, with $l=0.5\alpha=0.177$ $\mu$m, which is
very close to the tungsten particles ($r=0.169$ $\mu$m).
%
\begin{figure}[htb]
\includegraphics[height=0.9\columnwidth,angle=270]{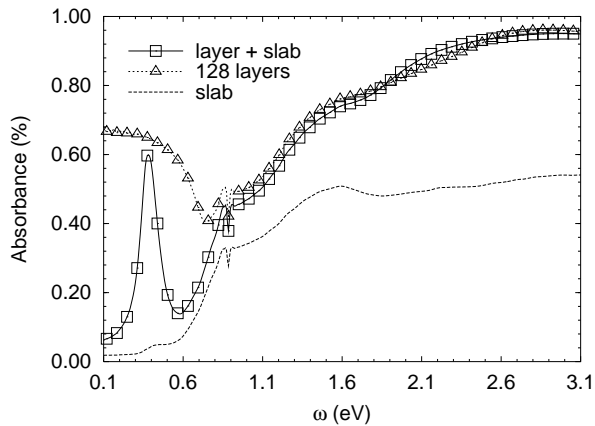}
\caption{Absorbance spectrum of a tungsten (111) layer put in front of a
tungsten slab (square), compared with the absorbance spectrum of a 128 layers
tungsten photonic crystal (upper triangle).  Light propagates along the
$\langle$111$\rangle$ direction and the interlayer distance $d$ between each
(111) layer is 0.289 $\mu$m.  The dielectric data of tungsten is taken from
Palik,\cite{palik_1985} and the absorbance spectrum of a uniform tungsten slab
with a thickness equal to 0.106 $\mu$m (dashed line) is shown as a reference.}
\label{w_layer_plus_w_slab}
\end{figure}
%
%
\begin{figure}[htb]
\includegraphics[height=0.9\columnwidth,angle=270]{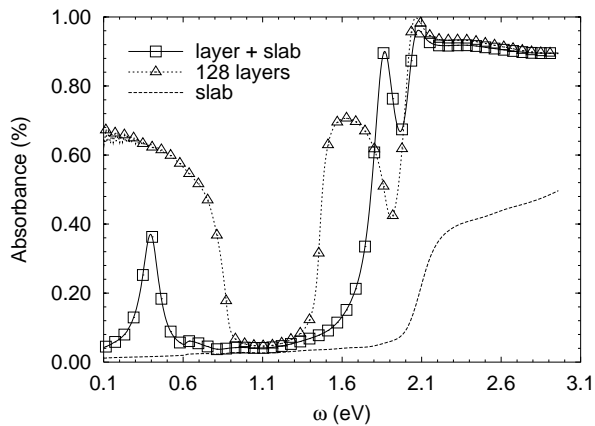}
\caption{Absorbance spectrum of the structures mentioned in
Fig.~\ref{w_layer_plus_w_slab}, but with tungsten replaced by copper.}
\label{cu_layer_plus_cu_slab}
\end{figure}
%
In Fig.~\ref{w_layer_plus_w_slab}, we plot the absorbance spectrum of the
proposed structure and compare it with the absorbance spectrum of a 128 layers
tungsten photonic crystal.  Again, the number of layers (128) is chosen such
that no light can be transmitted.  The spectra can be approximately separated
into a high-frequency ($\omega > 0.9$ eV) region and a low-frequency region
($\omega < 0.9$ eV).  At high-frequencies, the absorbance of the proposed
`layer + slab' (LS) surface ($A_{\rm LS}$) is close to that of the photonic
crystal ($A_{\rm PC}$) for $\omega > 0.9$ eV.  In particular, $A_{\rm LS}$ is
only 0.7\% and 1.4\% smaller than $A_{\rm PC}$ at 3.11 eV ($\lambda=0.4$
$\mu$m) and 1.78 eV ($\lambda=0.7$ $\mu$m) respectively.  Thus the present
micro-textured surface should be a good thermal emitter at optical
frequencies.  At low-frequencies $A_{\rm LS}$ is smaller than $A_{\rm PC}$ in
general.  It is interesting to note there is a sharp increase of $A_{\rm LS}$
between 0.3 eV and 0.5 eV. This is arising from the surface induced coupling
between the particles and the slab surface, which is a Fabry-Perot effect.
However in reality, it is
difficult to observe such coupling as the surfaces of the particles and the
slab are not perfect in general.  We also observe a suppression in $A_{\rm
PC}$ between 0.7 eV and 0.9 eV, which is due to photonic bandgap effect.  The
reason of choosing 0.9 eV as a benchmark is not only
phenomenological.  In the region 0.6 eV $< \omega <$ 0.9 eV, the magnitude of
the real part of the dielectric constant ($\epsilon_r(\omega)$) drops more
than 80\%, and decreases gradually after 0.9 eV, while its imaginary part only
has a small fluctuation.  The decrease of the dielectric mismatch between 0.6
eV and 0.9 eV allows more photons to penetrate into the material, which
results in the pronounced increase of $A_{\rm LS}$.  This can also be observed
for a tungsten slab ($A_{\rm slab}$).  In Fig.~\ref{cu_layer_plus_cu_slab}, we replace tungsten
by copper and plot the absorbance spectra. Again, there is a benchmark at
$\omega \approx$ 2.2 eV where $A_{\rm LS}$ is very
close to $A_{\rm PC}$ for $\omega >$ 2.2 eV and the drastic increase of the
absorbance for a copper slab stops accordingly. The physical reason of this is the
same as that of tungsten.

\begin{figure}[htb]
\includegraphics[height=0.9\columnwidth,angle=270]{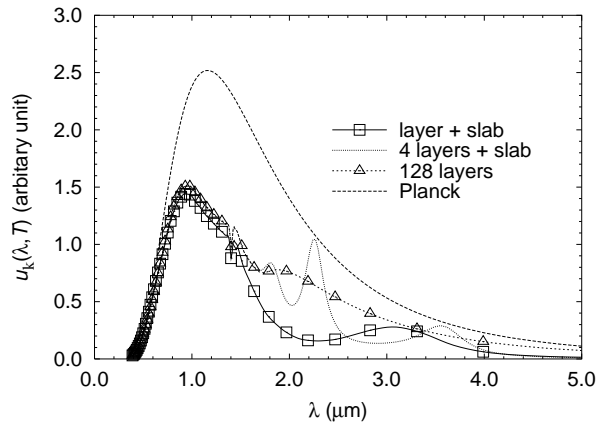}
\caption{Thermal emission spectra of the tungsten `layer + slab' surface,
`4 layers + slab' surface and
the 128 layers tungsten photonic crystal at 2500 K.  The emission peaks are
blue-shifted.  The corresponding Planck spectrum is also shown (dashed line).}
\label{thermal_spectra}
\end{figure}

We are now at a position to compare the thermal emission spectra between the
`layer + slab' surface and the 128 layers photonic crystal.  We will consider
the case of tungsten only because its melting point ($\approx$ 3700 K) is much higher
than that of copper ($\approx$ 1360 K), which makes tungsten a more feasible
material as a thermal emitter, the results are shown in
Fig.~\ref{thermal_spectra}.  The thermal emission spectrum is calculated by
assuming Kirchhoff's law,\cite{howell_2002} the thermal emission spectrum
$u_{\bf k}(\lambda,T)$ of wavevector ${\bf k}$ at temperature $T$ is given by:
\begin{equation}
u_{\bf k}(\lambda,T)=u_{b,{\bf k}}(\lambda,T)\times A_{\bf k}(\lambda),
\end{equation}
where $u_{b,{\bf k}}(\lambda,T)$ is the Planck spectrum of blackbody
radiation.
%
%
The temperature effect on the tungsten dielectric constant is important at
microwave and infrared frequencies.\cite{arnold_ao_1984,saxena_pcm_1997}
However, as our primary interest is to compare the differences between the
`layer + slab' surface and the photonic crystal, the study of the  temperature
effect is beyond the scope of this work and is thus ignored.  In
Fig.~\ref{thermal_spectra}, we show the thermal emission spectrum of the
`layer + slab' surface ($u_{\rm LS}$) at 2500 K and compare it with that of a
128 layers photonic crystal ($u_{\rm PC}$).  We also plot the spectrum of a `4
layers + slab' ($u_{\rm 4LS}$) to show the effects of extra layers.  In all
cases, the infrared part of the spectrum is effectively suppressed and the
optical spectrum is close to that of a blackbody radiation, thus the emission
peaks of both spectra are blue-shifted.  Additionally, the `layer + slab'
geometry emits less photons than that of the photonic crystal in the range
$1.2 <\lambda < 3.0$ $\mu$m, a direct consequence due to the small absorbance
[Fig.~\ref{w_layer_plus_w_slab}].  Next we compare the spectra $u_{\rm LS}$ and
$u_{\rm 4LS}$.  At short-wavelengths ($\lambda < 1.35$ $\mu$m), the addition
of extra layers only has an insignificant enhancement on the emission
spectrum.  In fact, $u_{\rm 4LS}$ almost coincides with $u_{\rm PC}$ in this
region.  In other words, only a few layers of tungsten particles are
sufficient for the purpose of optical emission, which is in accordance with
the results of Fleming et al.\cite{fleming_nature_2002}  We will have a
detailed discussion about the layer dependency in the next paragraph.

%
\begin{figure}[htb]
\includegraphics[height=0.9\columnwidth,angle=270]{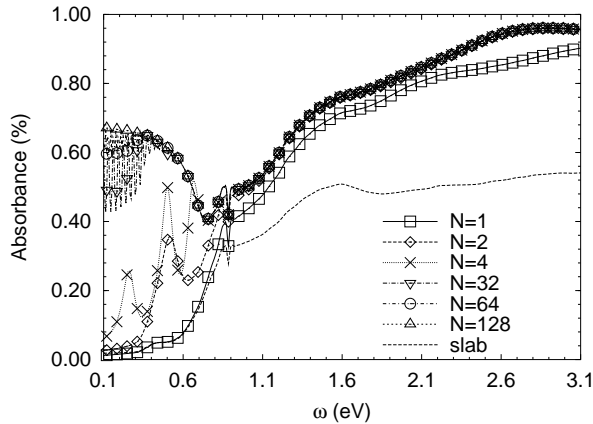}
\caption{Absorbance spectra of tungsten photonic crystals with different
numbers of (111) layers, $N = 1, 2, 4, 32, 64$ and 128 (various lines with
symbols).
The absorbance spectrum of a tungsten slab is also shown (dashed line).
The absorption is enhanced by the photonic crystals at all frequencies.}
\label{fcc_multi_layers}
\end{figure}
%

The results of Figs.~\ref{w_layer_plus_w_slab} and \ref{thermal_spectra}
suggest that, if compared with a 3D photonic crystal, a single tungsten
particle layer should be a better choice as a thermal emitter at optical
frequencies.  On one hand it absorbs more optical photons than a thick slab
does, and on the other hand the absorption of long-wavelength electromagnetic
waves is also suppressed.  It is thus instructive to examine the
layer-dependence of the absorbance spectrum.  In Fig.~\ref{fcc_multi_layers},
we show the absorbance spectra of tungsten photonic crystals with different
numbers of (111) layers ($N$).  As expected, each photonic crystal has its
absorbance exceeds that of a thick tungsten slab at all frequencies.  At
low-frequencies, we observe a $N$-dependency of the absorbance.  This
$N$-dependency stops until the number of layers is sufficiently large ($N\ge
128$), where all the photos penetrating into the photonic crystal are absorbed
by the tungsten particles.  We postpone the discussion of this $N$-dependency
until later.  We now concentrate on the high frequencies region, in this
region, a tungsten particle layer has a significant enhancement of absorbance
over that of a thick tungsten slab. Unlike the case of low-frequencies, the
absorbance is nearly $N$-independent for $N\ge 2$, indicating that the
attenuation is dominated by the first layer.  Physically speaking, if one
constructs a multi-layers photonic crystal, the second layer will further
absorb the photons penetrating the first layer, and the absorption will then
be saturated.  It explains the $N$-independency of absorbance for $N\ge 2$ in
Fig.~\ref{fcc_multi_layers}.  We then turn to the region $\omega < 0.9$ eV. In
this region, the absorbance of a tungsten particle layer is almost the same as
that of a tungsten slab. In the case of a slab, all the unabsorbed photons are
reflected and no transmission of photos is allowed, and absorbance spectrum
depends solely on the dielectric parameter of tungsten, while in the case of a
particle layer, photons can penetrate at all frequencies, owing to the fact
that the tungsten particles are not touching each others.  Thus in a photonic
crystal each layer will help absorbing the excess (long-wavelength)
electromagnetic waves scattered by previous layer.  This accounts for the
$N$-dependency of absorbance at low-frequencies.  At moderate frequencies
($0.7<\omega<0.9$ eV), the absorption is suppressed by photonic bandgap effect.

Here a few comments are in order.  The results of
Fig.~\ref{w_layer_plus_w_slab} and Fig.~\ref{thermal_spectra} suggest that, in
order to emit the same amount of light, less energy will be required to heat
the micro-textured surface, which is an advantage over the photonic crystal.
The fabrication of the tungsten particle layer can be done by cutting-age
techniques such as
self-assembly.\cite{velev_adv_mater_2000,zakhidov_synth_met_2001} Recently, Lu
et al.\cite{lu_nano_lett_2005} demonstrated a novel technique in fabricating
high-density silver nanoparticle layer.  The interparticle distance is tunable
and can be controlled precisely.  Their technique can be adopted to fabricate
a tungsten particle film.  It should also be emphasized that our discussion
here should be general for any 3D metallic photonic crystals.  If one
optimizes the geometric and materials, the suppression of absorption  below a
definite frequency can be accomplished by only a few (in our case, one)
building blocks of photonic crystals.  If the purpose is to suppress the
absorption within a certain frequency range, one has to utilize the photonic
bandgap effect and determine how many unit cells is necessary.

In this letter, we have proposed the use of a simple, micro-textured surface
as a thermal emitter.  We have studied its thermal emission behavior and
compare it with that of a tungsten photonic crystal.  The results suggest such
geometry would be a promising thermal emitter.  It is hoped that our study
would inspire new ideas in the design of thermal emitting applications.

This work is supported by RGC HK through Grant 600403.
We thank Jensen Li for discussions.


\begin{thebibliography}{10}

\bibitem{howell_2002}
R.~Siegel and J.~R. Howell,
\newblock {\em Thermal Radiation Heat Transfer},
\newblock Taylor and Francis, New York, 2002.

\bibitem{planck_1959}
M.~Planck,
\newblock {\em The Theory of Heat Radiation},
\newblock Dover, New York, 1959.

\bibitem{hesketh_nature_1986}
P.~J. Hesketh, J.~N. Zemel, and B.~Gebhart,
\newblock Nature {\bf 324}, 549 (1986).

\bibitem{joannopoulos_1995}
J.~D. Joannopoulos, R.~D. Meade, and J.~N. Winn,
\newblock {\em Photonic Crystals: Molding the Flow of Light},
\newblock Princeton Universiry Press, Princeton, N.J., 1995.

\bibitem{dowling_pra_1999}
C.~M. Cornelius and J.~P. Dowling,
\newblock Phys. Rev. A {\bf 59}, 4736 (1999).

\bibitem{pendry_jpcm_1999}
J.~B. Pendry,
\newblock J. Phys.: Condens. Mat. {\bf 11}, 6621 (1999).

\bibitem{chan_prl_2000}
{W. Y. Zhang et al.},
\newblock Phys. Rev. Lett. {\bf 84}, 2853 (2000).

\bibitem{chan_opt_express_2001}
W.~Y. Zhang, C.~T. Chan, and P.~Sheng,
\newblock Opt. Express {\bf 8}, 203 (2001).

\bibitem{fleming_nature_2002}
{J. G. Fleming et al.},
\newblock Nature {\bf 417}, 52 (2002).

\bibitem{modinos_cpc_1998}
N.~Stefanou, V.~Yannopapas, and A.~Modinos,
\newblock Comput. Phys. Commun. {\bf 113}, 49 (1998).

\bibitem{joannopoulos_prl_2004}
C.~Luo, A.~Narayanaswamy, G.~Chen, and J.~D. Joannopoulos,
\newblock Phys. Rev. Lett. {\bf 93}, 213905 (2004).

\bibitem{palik_1985}
E.~D. Palik, editor,
\newblock {\em Handbook of Optical Constants of Solids},
\newblock Academic Press, Orlando, 1985.

\bibitem{arnold_ao_1984}
G.~S. Arnold,
\newblock App. Opt. {\bf 23}, 1434 (1984).

\bibitem{saxena_pcm_1997}
L.~S. Dubrovinsky and S.~K. Saxena,
\newblock Phys. Chem. Minerals {\bf 24}, 547 (1997).

\bibitem{velev_adv_mater_2000}
O.~D. Velev and E.~W. Kaler,
\newblock Adv. Mater. {\bf 12}, 531 (2000).

\bibitem{zakhidov_synth_met_2001}
{A. A. Zakhidov et al.},
\newblock Synth. Met. {\bf 116}, 419 (2001).

\bibitem{lu_nano_lett_2005}
Y.~Lu, G.~L. Liu, and L.~P. Lee,
\newblock Nano Lett. {\bf 5}, 5 (2005).

\end{thebibliography}
\end{document}